\documentclass[twocolumn,showpacs,preprintnumbers,amsmath,amssymb,superscriptaddress]{revtex4}
\usepackage{epsfig,bm,color,ulem}

\begin{document}
\title{Glauber-model analysis of total reaction cross sections 
for Ne, Mg, Si, and S isotopes with Skyrme-Hartree-Fock densities}
\author{W. Horiuchi}
\affiliation{Department of Physics, Hokkaido University, Sapporo
060-0810, Japan}
\affiliation{RIKEN Nishina Center, Wako 351-0198, Japan}
\author{T. Inakura}
\author{T. Nakatsukasa}
\affiliation{RIKEN Nishina Center, Wako 351-0198, Japan}
\author{Y. Suzuki}
\affiliation{Department of Physics, Niigata University, Niigata
950-2181, Japan}
\affiliation{RIKEN Nishina Center, Wako 351-0198, Japan}
\pacs{21.10.Gv, 21.60.Jz, 25.60.Dz, 27.30.+t}

\begin{abstract}
A systematic analysis is made on the total reaction cross sections 
for Ne, Mg, Si, and S isotopes.
The high-energy nucleus-nucleus collision is described based on the Glauber model. 
Using the Skyrme-Hartree-Fock method in the three-dimensional 
grid-space representation, we determine the nuclear density 
distribution for a wide range of nuclei self-consistently 
without assuming any spatial symmetry. 
The calculated total reaction cross sections consistently agree 
with the recent cross section data on Ne$+^{12}$C collision 
at 240$A$\,MeV, which makes it possible to discuss 
the radius and deformation of the isotopes.
The total reaction cross sections for Mg$+^{12}$C,
Si$+^{12}$C and S$+^{12}$C cases are predicted for future measurements.
We also find that the high-energy cross 
section data for O, Ne, and Mg isotopes on a $^{12}$C target at around 1000\,$A$MeV 
can not be reproduced consistently with the corresponding data at 240\,$A$MeV.
\end{abstract}

\maketitle

\section{Introduction}

Advances in measurements of unstable nuclei 
have been providing information on exotic nuclei
toward the neutron and proton driplines.
Recently total reaction or interaction cross sections
have been measured in $pf$-shell region~\cite{kanungo11a, takechi10, takechi.new}.
They exhibit interesting trends which imply exotic structure, for example,
halos, skins, deformations etc., as we approach the driplines. 
In particular, the first observation of a halo structure 
of $^{31}$Ne~\cite{nakamura09} has stimulated 
several theoretical studies~\cite{horiuchi10, urata11, minomo12}.

Nuclear deformation is one of the unique properties of a
finite quantum system.
If the intrinsic wave function shows some deformation,  
the nuclear radius becomes effectively large
because the ground-state wave function
is expressed as a superposition of 
the intrinsic wave functions with different orientations.
Since the total and interaction cross sections 
are closely related to the nuclear radius,
it is interesting to investigate a relation between these cross sections
and nuclear deformation.

In this paper, we present a systematic analysis of total
reaction cross sections, on a $^{12}$C target, of unstable nuclei 
in the $sd$- and $pf$-shell region, focusing on
nuclear size properties, especially radius and deformation, 
and their relationship with the cross section. 
This study is motivated by the recent systematic measurements
of the total reaction cross sections 
of the $pf$-shell region~\cite{kanungo11a, takechi10, takechi.new}. 
We also predict the cross sections of other nuclei 
for future measurements.
The analysis will give us important information on the interplay between 
the nuclear structure and the cross section.

In Refs.~\cite{horiuchi07,ibrahim09}, two of the present authors 
(W.H. and Y.S.), B. Abu-Ibrahim, and A. Kohama 
performed systematic analyses of the total reaction
cross sections of C and O isotopes on a $^{12}$C target.
In these studies, the wave functions generated from 
a spherical Woods-Saxon model were
employed and the center-of-mass motion was appropriately removed.
The Glauber model, which gives a fair description of 
high-energy nucleus-nucleus collisions from nucleon degrees of freedom,
provided a good agreement with experiment.
However, in the present paper, we study the region including the island of inversion 
where the nuclear deformation may play a significant role.
Thus, we perform the Skyrme-Hartree-Fock calculation in the
three-dimensional (3D) coordinate-mesh representation which allows us
to treat the correct asymptotic behaviors of single-particle orbitals and 
the nuclear deformation self-consistently.
Although the parameter sets of Skyrme energy functionals are mostly
determined by fitting the properties of heavy closed-shell nuclei
and the nuclear matter,
they are known to give a good description of light nuclei as well.
This universality is one of the main advantages of the
density-dependent mean-field models.
In order to see the sensitivity of the nuclear deformation to the radius,
two different types of Skyrme parameter sets are employed for comparison.

This paper is organized as follows.
In Sec.~\ref{models.sec}, we briefly explain our reaction and structure models
to calculate the total reaction cross sections.
Sec.~\ref{results.sec} presents the calculated 
total reaction cross sections on a $^{12}$C target 
at the medium energy of 240 $A$MeV, corresponding to 
the recent experiments carried out at RIKEN.
We compare our theoretical cross sections and the experimental ones
for Ne in Sec.~\ref{Ne.result}.
Motivated by a very recent measurement~\cite{takechi.new}, 
we show our results for Mg isotopes in Sec.~\ref{Mg.result}.
Predictions for Si and S isotopes are also made in Sec.~\ref{Si_S.result}.  
We discuss in Sec.~\ref{problem} a problem concerning the total reaction cross section data at around 1000 $A$MeV for O, Ne, and Mg isotopes.  
Conclusion is drawn in Sec.~\ref{conclusions.sec}.

\section{Models}
\label{models.sec}

\subsection{Glauber model}

We describe a high-energy nucleus-nucleus collision
in the Glauber formalism~\cite{glauber}. 
The total reaction cross section is obtained by 
integrating the reaction probability $P(\bm{b})$ over the impact 
parameter $\bm{b}$
\begin{align}
\sigma_R=\int d\bm{b}\,P(\bm{b}),
\end{align}
with
\begin{align}
P(\bm{b})=1-|\text{e}^{i\chi(\bm{b})}|^2,
\label{reacprob.eq}
\end{align}
where $\chi(\bm{b})$ is the phase-shift function for the elastic 
scattering of a projectile ($P$) and a target ($T$). 
The phase shift function is given in terms of the ground-state 
wave functions of the projectile and the target as
\begin{align}
\text{e}^{i\chi(\bm{b})}&=\left<\Phi_0^P\Phi_0^T\right|\notag\\
&\times\prod_{i\in P}^{A_P}\prod_{j\in T}^{A_T}
\left[1-\Gamma_{NN}(\bm{s}_i^P-\bm{s}_j^T+\bm{b})\right]
\left|\Phi_0^P\Phi_0^T\right>,
\label{phasefn.eq}
\end{align}
where $\bm{s}_i$ is the transverse component of the $i$th nucleon coordinate
and $\Gamma_{NN}$ is the nucleon-nucleon profile function 
which we parametrize in the following form
\begin{align}
\Gamma_{NN}(\bm{b})=\frac{1-i\alpha_{NN}}{4\pi\beta_{NN}}\sigma_{NN}^\text{tot}
\exp\left(-\frac{\bm{b}^2}{2\beta_{NN}}\right).
\end{align}
The parameter set of $\sigma_{NN}^\text{tot}$, $\alpha_{NN}$, and 
$\beta_{NN}$ used here are
given in Ref.~\cite{ibrahim08}.

Though the multiple integration in Eq.~(\ref{phasefn.eq}) may be 
performed with a Monte Carlo integration as was done for light 
nuclei~\cite{varga02}, we take a usual approximation in this paper. 
The optical limit approximation (OLA) offers the most simple expression,
which is obtained 
by taking only the first-order term of 
the cumulant expansion~\cite{glauber} of Eq. (\ref{phasefn.eq}), that is
\begin{align}
\text{e}^{i\chi_\text{OLA}(\bm{b})}
&=\exp\left[-\iint d\bm{r}^Pd\bm{r}^T \rho_P(\bm{r}^P)\rho_T(\bm{r}^T)\right.
\notag\\
&\times\Gamma_{NN}(\bm{s}^P-\bm{s}^T+\bm{b})\bigg],
\end{align}
which involves a double-folding procedure of the projectile and target 
densities ($\rho_P$, $\rho_T$) with the effective interaction $\Gamma_{NN}$. 
The OLA misses some higher-order terms of $\Gamma_{NN}$ 
and multiple scattering effects. 
The resulting total reaction cross section of the nucleus-nucleus collision, 
tends to overestimate the measured cross section~\cite{horiuchi07}.

In Ref.~\cite{ibrahim00}, another expression of evaluating the 
phase shift function is proposed
in order to take account of the multiple scattering processes missing in the OLA.
To derive the formula, first we use
the cumulant expansion of the phase shift function 
for the nucleon-target system and 
take the first-order term back to the original phase shift function (\ref{phasefn.eq}).
Using the cumulant expansion again,
its leading term is called the Nucleon-Target formalism 
in the Glauber model (NTG) approximation, which is
\begin{align}
&\text{e}^{i\chi_\text{NTG}(\bm{b})}
=\exp\left\{-\int d\bm{r}^P\rho_P(\bm{r}^P)\right.
\notag\\
&\times\left.\left[1-\exp\left(-\int d\bm{r}^T
\rho_T(\bm{r}^T)\Gamma_{NN}
(\bm{s}^P-\bm{s}^T+\bm{b})\right)\right]\right\}.
\label{NTG.eq}
\end{align}
In numerical calculations shown in Sec.~\ref{results.sec},
we use the symmetrized expression of Eq. (\ref{NTG.eq}) 
by exchanging the role of the projectile and target nuclei.
Note that the NTG approximation
requires the same inputs as those of the OLA.
It gives a simple but fair description for high-energy reactions. For example,
the total reaction cross sections of $^{12}$C+$^{12}$C collisions are
improved very much in a wide energy range~\cite{horiuchi07}.

\subsection{Skyrme-Hartree-Fock method in 3D coordinate-mesh representation}
\label{SHF.sec}

We perform the Skyrme-Hartree-Fock calculation for the density
distribution of a variety of projectiles.
The ground state is obtained by minimizing the following energy density
functional\cite{VB72}
\begin{equation}
E[\rho] = E_N + E_C - E_{\rm cm} .
\end{equation}
For the ground states of even-even nuclei,
the nuclear energy $E_N$ is given by a functional
of the nucleon density $\rho_q(\bm{r})$, the kinetic density
$\tau_q(\bm{r})$, the spin-orbit-current density
$\bm\nabla\cdot\bm{J}_q(\bm{r})$ ($q=n,p$).
The Coulomb energy $E_C$ among protons is a sum of direct and
exchange parts.
The exchange part is approximated by means of the Slater approximation,
$\propto \int d\bm{r} \rho_p(\bm{r})^{4/3}$.
The center-of-mass recoil effect is also treated approximately
by a subtraction of the expectation value of
$E_{\rm cm}=(\sum_i p_i)^2/(2mA)\approx \sum_i p_i^2/(2mA)$.

Every single-particle wave function
$\phi_i(\bm{r},\sigma,q)$ is represented in the 3D grid points with
the mesh size of $\Delta x=\Delta y=\Delta z=$ 0.8 fm.
All the grid points inside the sphere of radius of 15 fm are
adopted in the model space.
The ground state is constructed by the imaginary-time method \cite{DFKW80}
with the constraints on the center-of-mass
and the principal axis:
\begin{equation}
\begin{split}
&\int d\bm{r} x \rho({\bm r})
=\int d\bm{r} y \rho({\bm r})
=\int d\bm{r} z \rho({\bm r})
 = 0 ,\\
&\int d\bm{r} xy \rho({\bm r}) = 
\int d\bm{r} yz \rho({\bm r}) = 
\int d\bm{r} zx \rho({\bm r}) = 0 .
\end{split}
\label{HFcm.eq}
\end{equation}
For odd-$A$ nuclei, we adopt the half-filling approximation, thus 
the time-odd densities, such as the current density and
the spin density, do not contribute to the present calculation.
The computer program employed in the present work has been developed
previously for linear-response calculations \cite{NY05,INY09,INY11},
including all the time-odd densities.
However, those time-odd parts automatically vanish
in converged self-consistent solutions.
Because of this simple filling treatment for odd-$A$ nuclei,
we do not expect a quantitative description of
odd-even effect in our calculation.
For nuclei very near the neutron drip line, the nuclear radius becomes
extremely sensitive to the neutron separation energy. In other words,
a light modification of the Skyrme parameter set could drastically
modify the final result. Thus, we do not show such results for the
drip-line nuclei in Sec.~\ref{results.sec}.

We calculate isotopes with even proton numbers,
$Z=8$, 10, 12, 14, and 16.
The two parameter sets of SkM* \cite{bartel82} and SLy4 \cite{Chan97}
are employed in the following, and
we compare results obtained with these two parameter sets.
The SkM* functional is known to well account for the 
properties of the nuclear deformation,
while the SLy4 is superior to SkM* in reproducing 
the total binding energy.

The self-consistent solution becomes, in most cases,
a deformed intrinsic state $|\Phi_K \rangle$ with
the definite $K$ quantum number.
The ground-state wave function in the laboratory frame is
constructed according to the strong-coupling scheme \cite{BM75} as
\begin{equation}
\begin{split}
|\Psi_{KIM}\rangle = \left(\frac{2I+1}{16\pi^2(1+\delta_{K0})}\right)^2
\left\{
  |\Phi_K\rangle D^I_{MK}(\omega) \right. \\
 \left. +(-1)^{I+K} |\Phi_{\bar{K}}\rangle D^I_{M-K}(\omega)
\right\}, 
\end{split}
\end{equation}
where $D^I_{MK}(\omega)$ is the $D$ function depending on the three Euler angles $\omega$.
The density distribution in the ground state with $I=K$ is
simply given by
\begin{equation}
\label{rho_SHF}
\rho_q^{\text{(Lab)}}(r) = \frac{1}{4\pi} \int d\hat{\bm{r}} \rho_q(\bm{r}) ,
\end{equation}
taking the average over states with different magnetic quantum number $M$.
The density $\rho_q^{\text{(Lab)}}(r)$ 
of Eq. (\ref{rho_SHF})  is utilized in Eq. (\ref{NTG.eq})
as the projectile density distribution $\rho_P(\bm{r})$. 
Since the Skyrme energy functional is constructed so as to give a
density distribution $\rho(r)$ in the center-of-mass frame, we
directly use the density in Eq. (\ref{rho_SHF}) in the NTG formula (\ref{NTG.eq}).
    
\section{Results and discussion}
\label{results.sec}

Here we treat a reaction on a $^{12}$C target.
Proton and neutron density distributions of $^{12}$C 
are obtained in the same way as in Ref.~\cite{horiuchi07}.
The obtained density reproduces
the proton charge radius determined by an electron scattering.
The total reaction cross sections of both $p$+$^{12}$C and  $^{12}$C+$^{12}$C are 
reproduced very well in a wide energy range~\cite{horiuchi07, ibrahim08}.
We adopt this density distribution for the target nucleus.
The density distribution of the projectile nucleus is determined
by the Skyrme-Hartree-Fock calculation as explained 
in Sec.~\ref{SHF.sec}.
For both the projectile and target, 
we use proton and neutron densities separately. 

\subsection{Neon isotopes}
\label{Ne.result}

\begin{figure}[ht]
% Figure 1
\begin{center}
\epsfig{file=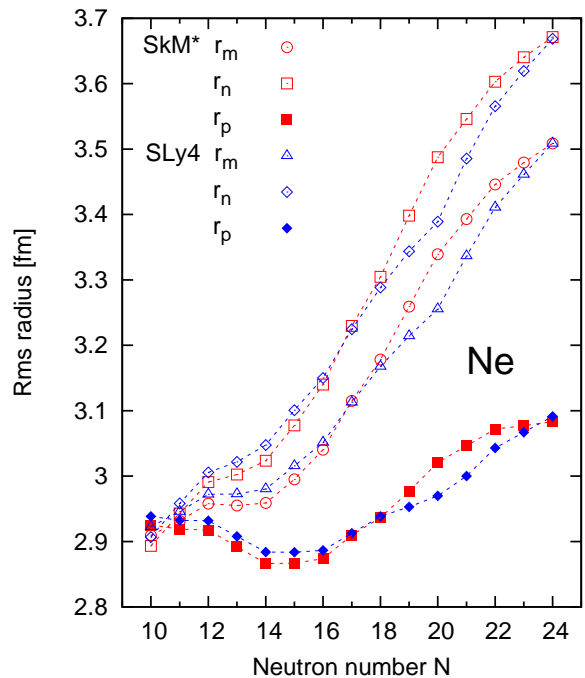,scale=1.3}
\caption{(Color online) Point matter, neutron, and proton rms 
radii of Ne isotopes calculated with the
SkM$^*$ and SLy4 interactions.}
\label{radiiNempn.fig}
\end{center}
\end{figure}

Figure~\ref{radiiNempn.fig} displays the point matter, neutron, and 
proton root-mean-square (rms) radii 
of Ne isotopes as a function of neutron number $N$. 
The enhancement of the matter radius in the neutron-rich isotopes
is dominantly due to that of the neutron radius.
An interesting observation is that both the neutron and proton radii follow 
the same behavior as the matter radius.
Both SkM$^*$ and SLy4 energy functionals produce similar results
for $N\lesssim 18$. 
We see, however, noticeable difference beyond $N=18$, and 
then two results coincide again at $N=24$.
What kind of structure property is responsible for these differences?

\begin{figure}[ht]
%Figure 2
\begin{center}
\epsfig{file=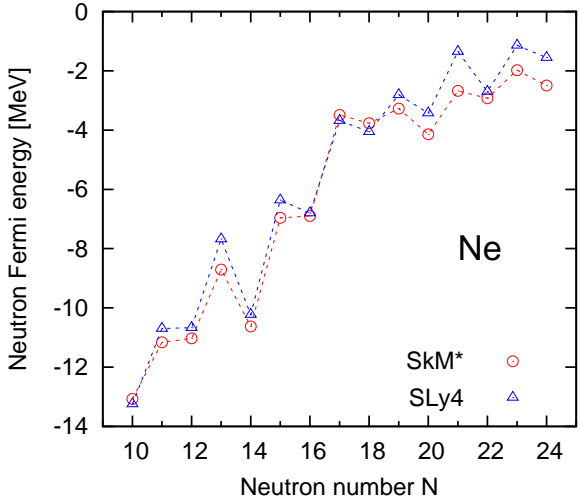,scale=1.3}
\caption{(Color online) Neutron Fermi energies of Ne isotopes 
as a function of neutron number $N$.}
\label{fermiNe.fig}
\end{center}
\end{figure}

Figure~\ref{fermiNe.fig} shows the calculated neutron Fermi energies.
Both SkM* and SLy4 results exhibit similar behavior, except for very neutron-rich
isotopes with $N>20$, in which
the neutron separation energy is calculated to be larger for SkM*
than SLy4.
Since a smaller separation energy is expected to produces a larger matter radius,
the behavior of the Fermi energy or the neutron separation energy does not explain the difference of the calculated matter radii
for nuclei with $18<N<24$ (Fig.~\ref{radiiNempn.fig}).
On the contrary, this gives an opposite effect.
The matter radius is predicted to be larger for the SkM*
calculation.

\begin{figure}[ht]
%Figure 3
\begin{center}
\epsfig{file=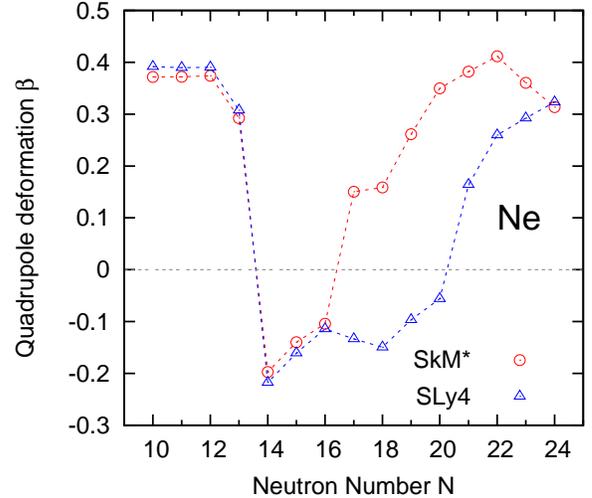,scale=1.3}
\caption{(Color online) Quadrupole deformation parameter $\beta$ of Ne isotopes calculated with the SkM$^*$ and SLy4 interactions.}
\label{betaNe.fig}
\end{center}
\end{figure}

The different matter radii are found to be mainly due to
the difference in the nuclear deformation.
We calculate the quadrupole deformation parameters from the mass quadrupole moments
\begin{align}
\beta&=\sqrt{\beta_{20}^2+\beta_{22}^2},
\end{align}
with
\begin{align}
\begin{split}
\beta_{20}&=\sqrt{\frac{\pi}{5}}\frac{\left<2z^2-x^2-y^2\right>}{\left<r^2\right>},\\
\beta_{22}&=\sqrt{\frac{3\pi}{5}}\frac{\left<y^2-x^2\right>}{\left<r^2\right>}.
\end{split}
\end{align}
The quantization axis is chosen as the largest (smallest) principal axis
for prolate (oblate) deformation.
Figure~\ref{betaNe.fig} presents the quadrupole deformation parameters $\beta$
obtained from the Hartree-Fock (HF) solutions.
The positive (negative) values of $\beta$ indicate the prolate (oblate)
deformations.
Beyond $N=16$, we find a significant difference in the magnitude of $\beta$,
between SkM* and SLy4.
In the SkM* calculation,
the deformation parameter $\beta$ is positive and rapidly increases for
$N>18$.
On the other hand, the SLy4 predicts the nearly spherical ground states
(weakly oblate) for $N\lesssim 20$ and changes its shape into prolate
for $N>20$. The value of 
$|\beta|$ is always larger in SkM* for $18<N<24$.
This nicely corresponds to the observation of the different behavior
in the matter radii at $18<N<24$ in Fig.~\ref{radiiNempn.fig}.
The matter radii coincide again at $N=24$,
so does the predicted $\beta$ values.
It should be noted that the kink behavior in the matter radii at $N=20$  (SLy4)
in Fig.~\ref{radiiNempn.fig} can be explained by the onset of the deformation at $N>20$,
while the kink at $N=14$ is due to the occupation of the $s_{1/2}$
orbitals at $N>14$. 

The proton number $Z=10$ of Ne isotope is known to strongly favor
the prolate shape, because it corresponds to the magic number at
the superdeformation~\cite{NY05,BM75}.
This is consistent with the results in Fig. \ref{betaNe.fig}
for Ne isotopes with $N\sim 10$.
Recent experiments reveal that the excitation energy of the first
$J^\pi=2^+$ states $E(2_1^+)$
in even Ne isotopes decrease from $N=16$ to $N=22$ \cite{Door09}.
This is qualitatively consistent with the increase of the quadrupole
deformation calculated with SkM*.

\begin{figure}[ht]
%Figure 4
\begin{center}
\epsfig{file=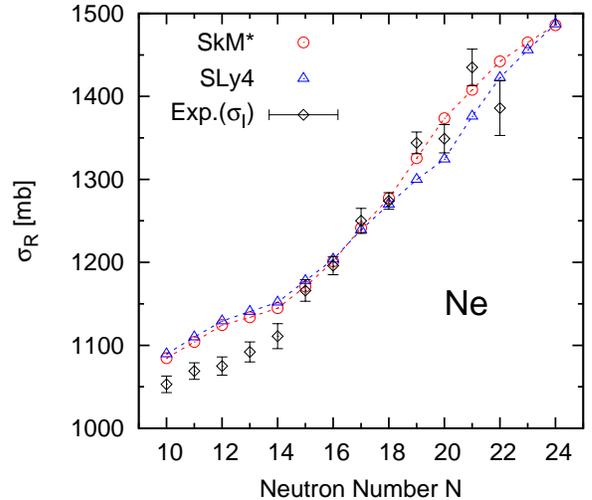,scale=1.3}
\caption{(Color online) 
Total reaction cross sections of Ne isotopes on a 
$^{12}$C target at 240 $A$MeV.
Experimental data are the interaction cross sections 
at 240 $A$MeV taken from Ref.~\cite{takechi10}.}
\label{reacNe.fig}
\end{center}
\end{figure}

The calculated total reaction cross sections of Ne isotopes
incident on a $^{12}$C target at 240 $A$ MeV 
are shown in Fig.~\ref{reacNe.fig}, as a function of neutron number $N$.
The agreement with the recent data \cite{takechi10} is very good.
The slope of the curve in Fig.~\ref{reacNe.fig} changes at $N=14$.
The cross section increases gradually from $N=10$ to 14
and shows a rapid rise from $N=14$.
This neutron-number dependence of the total reaction cross sections
also well corresponds to the calculated matter radii.

For the nuclei with $N=10\sim14$,
the calculated cross sections are slightly larger than the observed values,
while the agreement between the theory and experiment is fairly good 
in the neutron-rich region beyond $N=14$.
This may be explained by the difference between the total reaction 
and interaction cross sections. 
The theory gives 
the total reaction cross section $\sigma_R$ while the experiment is  the interaction cross section $\sigma_I$. 
Since the latter cross section does not contain inelastic reaction cross sections,
the inequality $\sigma_R \gtrsim \sigma_I$ always holds.
Contributions of the inelastic processes are classified as two cases:
(i) The projectile nucleus is excited to particle-bound excited states 
that are located below particle-emission threshold, while the target can take any states. 
(ii) The target nucleus is excited, but the projectile remains in 
the ground state. 
The difference between the total reaction and interaction 
cross sections is expected to be smaller as going closer to the neutron drip line. The difference is estimated to be at most 100 mb 
in a phenomenological way~\cite{kohama08}. 

For the neutron-rich nuclei $^{28-32}$Ne,
the results of SkM$^*$ nicely reproduce an average behavior
of the experimental cross sections,
but not their odd-even staggering.
This may be due to the pairing correlation \cite{HS12},
which is neglected in the present study.
Our result for these nuclei is similar to the one obtained
with the deformed Woods-Saxon potential \cite{minomo11}.
Approaching the neutron dripline,
the neutron separation energy becomes very small.
In such a situation, a spatial extension of the last neutron orbit
is very sensitive to the separation energy.
The 3D coordinate-space representation for the single-particle orbitals 
is suitable for the treatment of their asymptotics.
However, in the case of $^{31}$Ne,
the measured neutron separation energy is only about 0.33 MeV with
a large error bar~\cite{audi03},
while the calculated separation energy is as large as 2 MeV
with the SkM$^*$ interactions.
Together with the half-filling approximation,
the present Skyrme functionals may not provide us with such a precise
description of a loosely bound system.
To artificially adjust the Fermi level to 300 keV,
we multiply the mean-field potentials
($U_n$ and $U_p$ \cite{BFH87}) obtained with SkM$^*$
by a common scaling factor, then,
the total reaction cross section is easily increased by 140 mb.
This, in fact, significantly overestimates the experimental cross section.
Thus, an accurate measurement of the neutron separation energy is
highly desired.

\begin{figure}[ht]
%Figure 5
\begin{center}
\epsfig{file=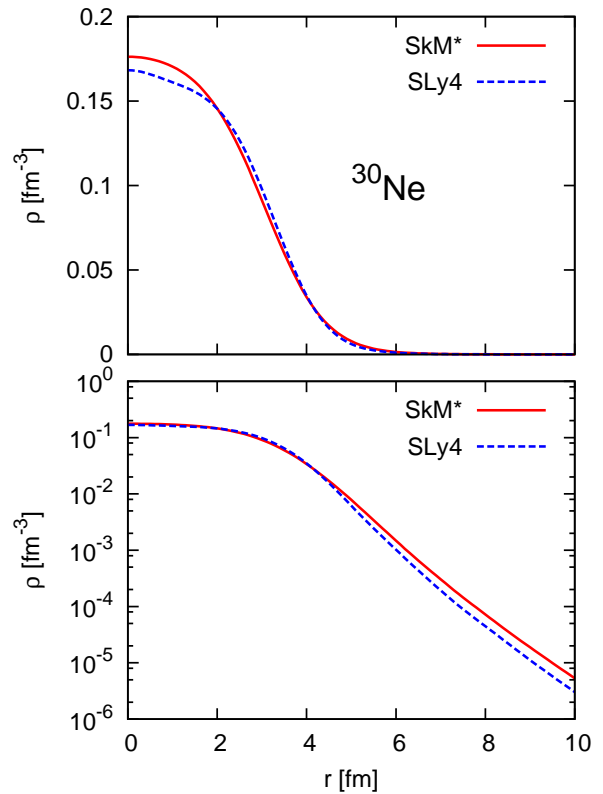,scale=1.3}
\caption{(Color online) Upper panel: Point matter density distributions of $^{30}$Ne 
calculated with the SkM$^*$ and SLy4 interactions.
Lower panel: The same as the upper panel but drawn in a logarithmic scale.}
\label{densNe.fig}
\end{center}
\end{figure}

Figure~\ref{densNe.fig} presents the calculated density profiles of $^{30}$Ne
in both linear and logarithmic scales.
The nucleus $^{30}$Ne is strongly deformed 
with SkM$^*$ ($\beta=0.35$), 
while the shape is almost spherical with SLy4 ($\beta=-0.06$).
In this case, two parameter sets exhibit quite different
density profiles, especially in the outer region that is crucial 
to determine the matter radius.
The density calculated with SkM$^*$ is larger than the one 
with SLy4 at $r\gtrsim 5$ fm.

\begin{figure}[ht]
%Figure 6
\begin{center}
\epsfig{file=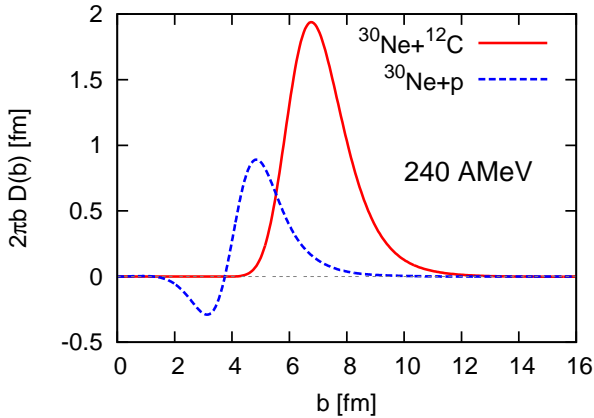,scale=1.3}
\caption{(Color online) Impact parameter dependence of the difference of the reaction probabilities 
for $^{30}$Ne calculated with the SkM* and SLy4 interactions.}
\label{reacprobNe.fig}
\end{center}
\end{figure}

In order to see how the above density 
difference leads to the difference in the total reaction cross 
section, we plot the difference of 
the reaction probabilities calculated with the SkM$^*$ and SLy4 
interactions 
\begin{align}
D(\bm{b})=P_{\text{SkM*}}(\bm{b})-P_{\text{SLy4}}(\bm{b}).
\end{align} 
Note that $D(\bm{b})$ depends only on $b=|\bm{b}|$. 
In Fig.~\ref{reacprobNe.fig}, 
we show $D(b)$ multiplied by $2\pi b$ for $^{30}$Ne projectile on 
both $^{12}$C and proton targets for comparison.
Though we see some difference in the interior density as shown 
in Fig.~\ref{densNe.fig}, it has almost no contribution to 
$D(b)$ for a $^{12}$C target, 
while $D(b)$ on a proton target 
shows oscillatory behavior reflecting
the difference of the density profiles between 2$\sim$4 fm.
The reaction on a $^{12}$C target 
is insensitive to the internal density profiles.
In contrast, we can see the difference in the outer region 
(Fig.~\ref{reacprobNe.fig}).
The reaction on a $^{12}$C target is advantageous to probe
the density at large distances. 

\subsection{Magnesium isotopes}
\label{Mg.result}

\begin{figure}[ht]
%Figure 7
\begin{center}
\epsfig{file=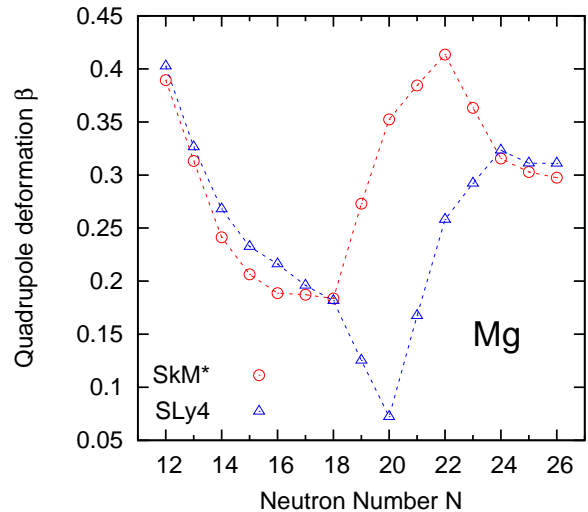,scale=1.3}
\caption{(Color online)  
Quadrupole deformation parameter $\beta$ of Mg isotopes 
calculated with the SkM$^*$ and SLy4 interactions.}
\label{betaMg.fig}
\end{center}
\end{figure}

\begin{figure}[ht]
%Figure 8
\begin{center}
\epsfig{file=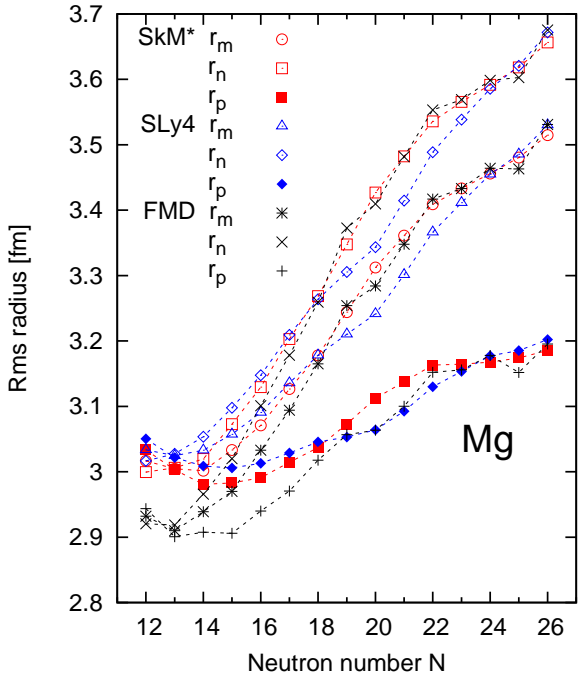,scale=1.3}
\caption{(Color online) 
Point matter, neutron, and proton rms radii of Mg isotopes calculated with the SkM$^*$ and SLy4 interactions. The radii obtained by the FMD~\cite{yordanov12,neff12} are also shown.}
\label{radiiMg.fig}
\end{center}
\end{figure}

Let us next discuss Mg isotopes.
Figures~\ref{betaMg.fig} and ~\ref{radiiMg.fig}
display the quadrupole deformation parameters
and the matter, neutron, and proton rms radii
for Mg isotopes.
The trend of the deformation parameters
is very similar to that of Ne isotopes.
Similarly to Ne case, 
the $\beta$ values obtained with SkM* 
are larger than that with SLy4 for $18<N<24$. 
This behavior influences the matter radii as well as the
neutron and proton radii as shown in Fig. \ref{radiiMg.fig}.
We also plot the radii obtained by the
fermionic molecular dynamics approach (FMD)~\cite{yordanov12,neff12}.
Though the FMD radii tend to be slightly smaller than our HF radii
in $12\leq N<18$, the HF radii with SkM* are very close to the FMD radii for
$N>18$. As shown in Ref.~\cite{yordanov12},
the trend of the quadrupole deformations 
is also similar to that obtained with SkM*.

The deformation $\beta$ calculated with SkM* is very similar to that
of Ne case.
The difference can be observed only for $N=14\sim 16$: 
the Ne isotopes have oblate shape with small deformation, while 
the Mg isotopes have prolate shape. 
Experimental excitation energies of the first $2^+$ states, 
$E(2_1^+)$, of $^{28,30}$Mg ($N=16$, 18)
are almost the same, whereas we see 
a rapid decrease of $E(2_1^+)$ in $^{26,28}$Ne~\cite{Door09}. 
Again, this is qualitatively consistent with the calculation.

\begin{figure}[ht]
%Figure 9
\begin{center}
\epsfig{file=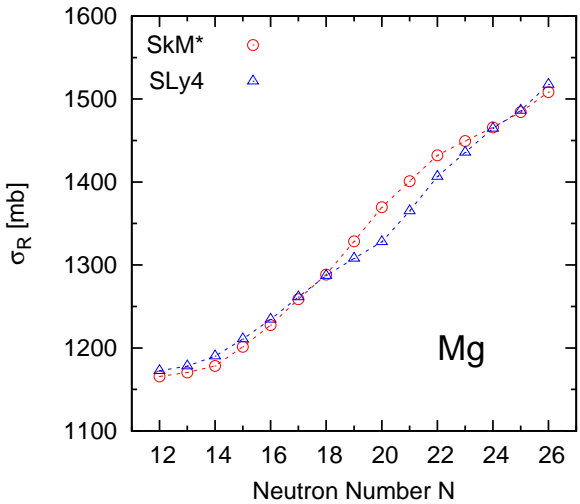,scale=1.3}
\caption{(Color online) 
Total reaction cross sections of Mg 
isotopes incident on a $^{12}$C target at 240 $A$MeV. 
}
\label{reacMg.fig}
\end{center}
\end{figure}

In Fig.~\ref{reacMg.fig} we show the total reaction cross sections
calculated using the SkM* and SLy4 interactions
for Mg isotopes incident on a $^{12}$C target.
A comparison with experiment is very interesting to judge
whether the neutron rich Mg isotopes
favor the strong deformation or not 
and also whether the HF densities at $12\leq N <18$ can reproduce
experimental cross sections or not because they predict
larger radii than the FMD calculation. 
Here we comment on the cross section of $^{37}$Mg.
Since the neutron separation energy of $^{37}$Mg is considered to be
small ($\sim$162 keV)~\cite{cameron12}, and the $\beta$ value of $^{37}$Mg is fairly large, a situation similar to $^{31}$Ne may occur in $^{37}$Mg as well.
If that is the case, the cross section predicted for $^{37}$Mg will be
increased than the present value.
We hope the measured values of the total reaction (interaction) cross sections 
are made available soon~\cite{takechi.new}.

\subsection{Silicon and sulfur isotopes}
\label{Si_S.result}

\begin{figure}[ht]
%Figure 10
\begin{center}
\epsfig{file=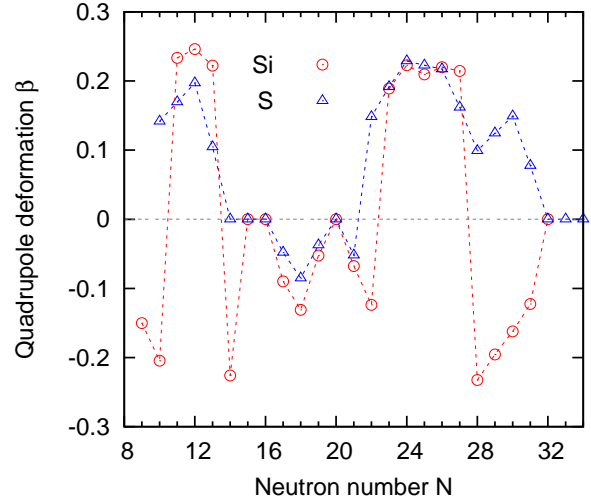,scale=1.3}
\caption{(Color online) 
Quadrupole deformation parameters of Si and S isotopes 
calculated with the SkM$^*$ interaction as a function of 
neutron number $N$.}
\label{betaSiS.fig}
\end{center}
\end{figure}

\begin{figure}[ht]
%Figure 11
\begin{center}
\epsfig{file=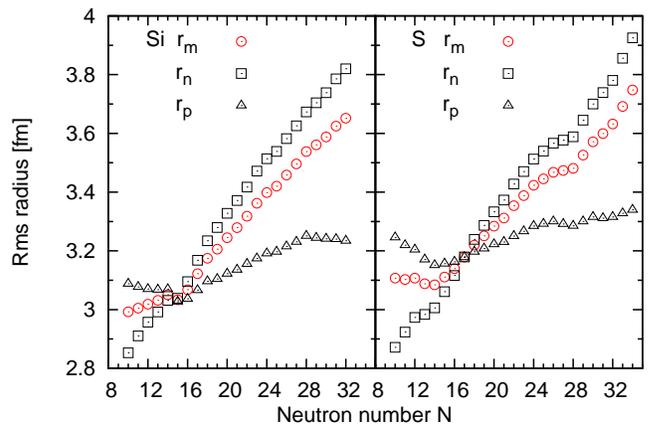,scale=1.15}
\caption{(Color online) Point matter, neutron, and proton rms radii
of Si (left) and S (right) isotopes as a function of neutron number $N$. The SkM$^*$ interaction is used.}
\label{radiiSiS.fig}
\end{center}
\end{figure}

We now discuss the cases of Si and S isotopes,  which will show deformation different from Ne and Mg isotopes.
Figures~\ref{betaSiS.fig} and \ref{radiiSiS.fig} plot the quadrupole deformation parameters as well as 
the point matter, neutron, and proton rms radii for Si and S isotopes, respectively.
We also see some kinks in the calculated matter, neutron, and proton radii
in Fig.~\ref{radiiSiS.fig}, 
corresponding to the peaks of the quadruple deformation
parameter displayed in Fig.~\ref{betaSiS.fig}.
The magnitudes of $\beta$ for Si and S isotopes are significantly smaller than those 
of Ne and Mg isotopes, in the region of $N\approx 20$, 
which suggests that the island of inversion does not reach $Z=14$ (Si).
This is probably due to that 14 protons favor  
the oblate shape, while the neutrons tend to deform the system 
into a prolate shape for $N>16$.
Thus, the competition between protons and neutrons determines the
deformation. 
The enhancement of the matter radii for $N>14$ is not as strong as
that in Ne and Mg isotopes, corresponding to the moderate change of $\beta$ values.
The matter radii of Si and S isotopes show similar enhancement.
The proton radii of S isotopes, however, tend to be larger than 
those of Si isotopes due to the occupation of the $s$ orbit. 
The neutron and matter radii of S isotopes 
also show kink behaviors at $N=28$.
This may be due to the decrease of the deformation at $N=26$ to 28
and the occupation of low-$\ell$ ($p$) orbitals
beyond $N=28$. Note that similar kink behaviors at $N=28$ are predicted in
the low-energy $E1$ strength for S isotopes but not for Si isotopes~\cite{INY11}.

\begin{figure}[ht]
%Figure 12
\begin{center}
\epsfig{file=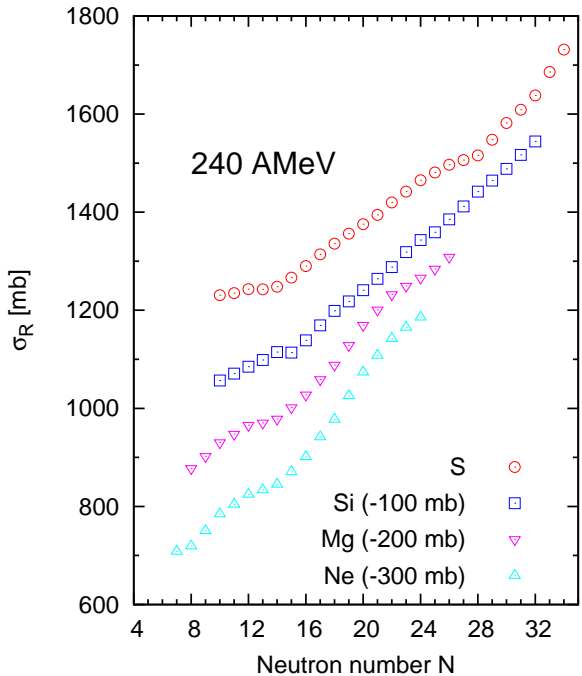,scale=1.3}
\caption{(Color online) Total reaction cross sections of Ne, Mg, Si 
and S isotopes on a $^{12}$C target at 240 $A$MeV as a function of neutron number $N$. The SkM$^*$ interaction is used. 
The cross sections for Ne, Mg, and Si isotopes are plotted by subtracting 300, 200, and 100\,mb, respectively.}
\label{rcs.fig}
\end{center}
\end{figure}

In Fig.~\ref{rcs.fig}, we plot the predicted total reaction cross sections 
of Si and S isotopes on a $^{12}$C target at 240 $A$ MeV. For the sake of 
comparison, the $\sigma_R$ values for Ne and Mg isotopes are also drawn.
Though some behavior indicating the nuclear deformation 
appears to persist in Si and S isotopes as well, 
the change of the cross sections in $18\leq N\leq 24$
is not as drastic as that of the cross sections of Ne and Mg isotopes.
In fact, the increase of $\sigma_R$ from $N=18$ to 24
is 210, 180, 140, and 130 mb, for Ne, Mg, Si, and S isotopes, respectively.

\subsection{Problems in the high-energy data}
\label{problem}

We have shown that the present theory reproduces 
the total reaction or interaction cross section data at the medium energy of 240\,$A$MeV. 
Since it is based on the Glauber model, 
our theory should provide us with a very good description of high-energy collisions 
around 1000\,$A$MeV. In fact, as shown in Ref.~\cite{horiuchi07}, 
the total reaction cross section of $^{12}$C+$^{12}$C collision is  
reproduced very well in the wide energy ranging up to 1000\,$A$MeV: 
The calculated cross sections at 240 and 1000\,$A$MeV are 790  and 850\,mb, respectively, which 
is in excellent agreement with the measured total reaction (interaction) 
cross sections, 782$\pm 10$\,mb~\cite{takechi09} at 250\,$A$MeV and 
853$\pm 6$\,mb~\cite{ozawa01} at 950\,$A$MeV.
We point out in this subsection, however, that some cross sections 
for O, Ne, and Mg isotopes at around 1000\,$A$MeV are difficult to reproduce.

\begin{figure}[ht]
%Figure 13
\begin{center}
\epsfig{file=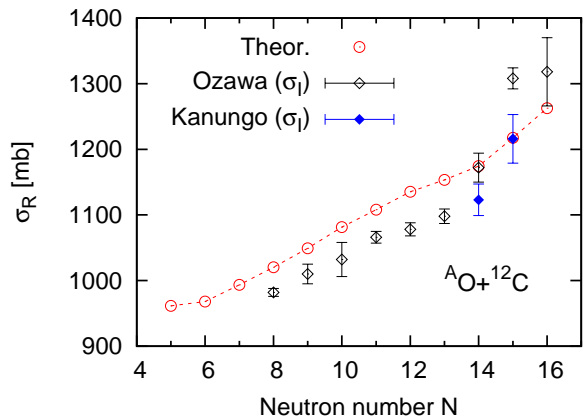,scale=1.3}
\caption{(Color online) 
Total reaction cross sections of O isotopes on a 
$^{12}$C target at 1000 $A$MeV. 
The SkM$^*$ interaction is used.  
Measured interaction cross sections
at around 950 $A$MeV are taken from Ref.~\cite{ozawa01} (open diamonds).
Most recently measured interaction cross sections by Ref.~\cite{kanungo11b} 
are denoted by filled diamonds.}
\label{reacO.fig}
\end{center}
\end{figure}

We start with a problem concerning the interaction cross sections of O isotopes.
The anomalously large cross section of $^{23}$O on a $^{12}$C target at around 1000$A$\,MeV has been 
a long standing problem~\cite{ozawa01,kanungo01,kanungo02}. 
A large jump of the measured  cross section $\sigma_I$ from $^{22}$O to $^{23}$O was analyzed by a model 
of adding one neutron in a loosely bound
$1s$ orbit to the $^{22}$O core~\cite{ozawa00}.
However, it failed to explain the data because one neutron separation energy 
(2.74 MeV) is too large to form a halo structure.
Figure~\ref{reacO.fig} displays the total reaction cross sections 
of O isotopes calculated with the SkM$^*$ interaction. 
Measured interaction cross sections are taken from 
Ref.~\cite{ozawa01} for $^{13-24}$O and from Ref.~\cite{kanungo11b} for $^{22,23}$O. 
The latter are new data remeasured in 2011. In the case of 
$^{23}$O, our result agrees well with the new data~\cite{kanungo11b} as well as the calculation based on a phenomenological Woods-Saxon potential~\cite{ibrahim09}.
The former experimental value for $^{23}$O~\cite{ozawa01} 
seems to be too large. 

\begin{figure}[ht]
%Figure 14
\begin{center}
\epsfig{file=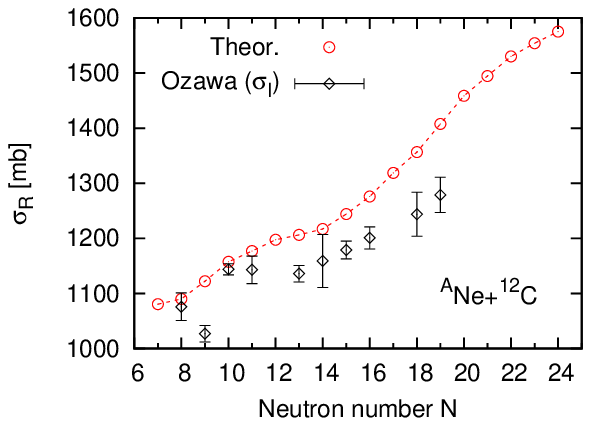,scale=1.3}
\epsfig{file=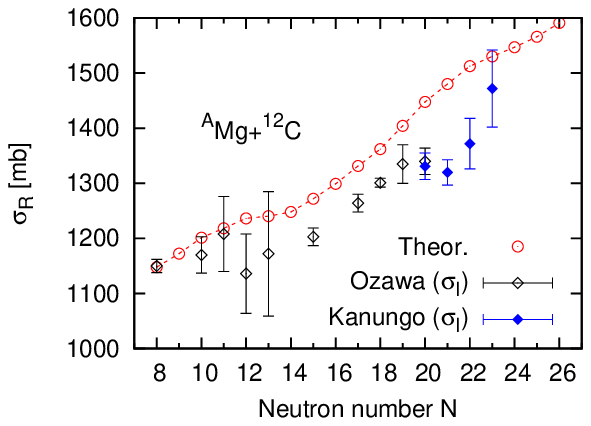,scale=1.3}
\caption{(Color online) Upper panel: Total reaction cross sections of Ne isotopes on a $^{12}$C target
at 1000\,$A$MeV. The SkM$^*$ interaction is used. Measured interaction cross sections at around 950\,$A$MeV
are taken from Ref.~\cite{ozawa01}.
Lower panel: 
Total reaction cross sections of Mg isotopes 
on a carbon target at 900\,$A$MeV. The SkM$^*$ interaction is used. 
Measured interaction cross sections
at around 950\,$A$MeV are taken from Refs.~\cite{ozawa01} (closed diamonds)
and~\cite{kanungo11a} (filled diamonds).}
\label{reacNehigh.fig}
\end{center}
\end{figure}

Next, we show in Fig.~\ref{reacNehigh.fig} the total reaction cross sections of Ne and Mg isotopes
on a $^{12}$C target at around 1000\,$A$MeV.
Experimental cross sections are taken 
from Refs.~\cite{ozawa01,kanungo11a}.
Again,
a considerable difference between the theory and experiment 
is observed. This difference is unexpected, however,  
in view of the fact that we have a good agreement with 
the measurement at 240\,$A$MeV for Ne isotopes (Fig.~\ref{reacNe.fig}). 
It is also noted 
that, although the calculation produces a smooth neutron number
dependence, the measured cross sections show an irregular
decrease at $^{33}$Mg. 

Since our calculations at 240\,$A$MeV 
reproduce the recent measurement very well and 
since the total reaction cross sections are expected
to be very close to the interaction cross sections 
at around 1000\,$A$MeV, we think that updating 
experimental cross section data is necessary to resolve the above problems.

\section{Conclusion}
\label{conclusions.sec}

We have made a systematic analysis of the total reaction cross sections of 
Ne, Mg, Si, and S isotopes on a $^{12}$C target.
The densities are obtained using the Skyrme-Hartree-Fock method
on a full three-dimensional grid space.
High-energy nucleus-nucleus collisions are described using the Glauber model.
Both structure and reaction models employed here have no adjustable parameter.

Comparing our results with the recent measurements 
of the total reaction cross sections for
Ne isotopes at 240\,$A$MeV, we find that a good 
agreement is obtained for both stable and unstable projectiles. 
We have shown that the nuclear deformation plays 
an important role to determine
the matter radius of the neutron-rich isotopes.
A similar trend in the total reaction cross sections
is expected for Mg isotopes whose deformation behaves like that of Ne isotopes.
We have also shown that 
the total reaction cross sections on a $^{12}$C target 
is sensitive to the nuclear density, especially 
near its surface where the deformation contributes 
to changing the density profile.
The systematic measurements of $\sigma_R$ ($\sigma_I$)
for a long chain of isotopes may reveal the enhancement of the nuclear size
which can be a signature of the nuclear deformation.
We have found that the proton radii also follow
the same behavior of the matter radii.
If one can measure the proton radii, for example,
by charge changing cross sections, they also give information
on nuclear deformations as well as nuclear skin-thicknesses.

We have predicted the total reaction cross sections for Mg, 
Si and S isotopes on a $^{12}$C target. 
We have also pointed out some contradictions 
in the high-energy cross section data at around 1000\,$A$MeV.  
In spite of the fact that we can excellently reproduce the experimental data at 240\,$A$MeV, we observe considerable disagreement between our calculation and the measured cross sections for O, Ne, and Mg isotopes on a $^{12}$C target at around 1000\,$A$MeV. Resolution of these problems requires further investigations 
in both theory and experiment.

It is interesting to examine the total reaction cross sections 
on a different target because they give different sensitivity 
to the density profile of the projectile nucleus. 
Since a $^{12}$C target has a finite size,
the reaction mainly takes place at the nuclear surface
where more particles are involved.
On the contrary, a proton target, as a point particle, is able to probe the inner region of the nucleus.
This will provide us with 
information on the interior density differently from the carbon target.
The work along this direction is underway.

\section*{Acknowledgments}

The authors thank M. Takechi and M. Fukuda for
valuable communication.
We also thank T. Neff for making his FMD results on the radii of 
Mg isotopes available to us.
This work is supported by the Grant-in-Aid for Scientific Research on 
Innovative Areas (No. 20105003), by the Grant-in-Aid for Scientific 
Research(B) (No. 21340073), and
by the Grants-in-Aid for Scientific Research(C) (Nos. 21540261 and 24540261). 
Most works of W.H. were done while he was supported by the Special Postdoctoral Researcher 
Program of RIKEN (2011.4-2012.4). 
T.I. is supported by the Special Postdoctoral Researcher Program of RIKEN.


\begin{thebibliography}{99}
\bibitem{kanungo11a} R. Kanungo {\it et al.}, Phys. Rev. C {\bf 83}, 021302 (R) (2011).
\bibitem{takechi10} M. Takechi {\it et al.}, Mod. Phys. Lett. A {\bf 25}, 1878 (2010).
\bibitem{takechi.new} M. Takechi, private communication; M. Fukuda 25aXA-8, JPS annual meeting (Nishinomiya, 2012.3).
\bibitem{nakamura09} T. Nakamura {\it et al.}, Phys. Rev. Lett. {\bf 103}, 262501 (2009).
\bibitem{horiuchi10}  W. Horiuchi, Y. Suzuki, P. Capel, D. Baye, Phys. Rev. C {\bf 
81}, 024606 (2010).
\bibitem{minomo12} K. Minomo, T. Sumi, M. Kimura, K. Ogata, Y. R. Shimizu, and M. Yahiro, Phys. Rev. Lett. {\bf 108}, 052503 (2012).
\bibitem{urata11} Y. Urata, K. Hagino, and H. Sagawa,
Phys. Rev. C {\bf 83}, 041303(R) (2011).
\bibitem{horiuchi07} W. Horiuchi, Y. Suzuki, B. Abu-Ibrahim, and A. Kohama, Phys. Rev. C {\bf 75}, 044607 (2007).
\bibitem{ibrahim09} B. Abu-Ibrahim, S. Iwasaki, W. Horiuchi, A. Kohama, and Y. Suzuki, J. Phys. Soc. Jap., Vol. 78, 044201 (2009).
\bibitem{glauber} R. J. Glauber, {\it Lecture in Theoretical Physics}, edited by W. E. Brittin and L. G. Dunham (interscience, New York, 1959), Vol. 1, p.315.
\bibitem{ibrahim08} B. Abu-Ibrahim, W. Horiuchi, A. Kohama, and Y. Suzuki, Phys. Rev. C {\bf 77}, 034607 (2008).
\bibitem{varga02} K. Varga, S.C. Pieper, Y. Suzuki, R.B. Wiringa, Phys. Rev. C {\bf 66}, 034611 (2002).
\bibitem{ibrahim00} B. Abu-Ibrahim and Y. Suzuki, Phys. Rev. C {\bf 62}, 034608 (2000).
\bibitem{VB72} D. Vautherin and D.M. Brink, Phys. Rev. C {\bf 5}, 626 (1972).
\bibitem{DFKW80} K.~T.~R. Davies, H. Flocard, S. Krieger, and M.~S. Weiss,
Nucl. Phys. A {\bf 342}, 111 (1980).
\bibitem{NY05} T. Nakatsukasa and K. Yabana,
Phys. Rev. C {\bf 71}, 024301 (2005).
\bibitem{INY09} T. Inakura, T. Nakatsukasa, and K. Yabana,
Phys. Rev. C {\bf 80}, 044301 (2009).
\bibitem{INY11} T. Inakura, T. Nakatsukasa, and K. Yabana,
Phys. Rev. C {\bf 84}, 021302 (2011).
\bibitem{bartel82} J. Bartel {\it et al.}, Nucl. Phys. {\bf A386}, 79 (1982).
\bibitem{Chan97}E. Chanbanat, P. Bonche, P. Haensel, J. Mayer, and R. Schaeffer,
Nucl. Phys. A {\bf 627}, 710 (1997).
\bibitem{BM75}A. Bohr and B. R. Mottelson,
{\it Nuclear Structure} Vol. 2 (W. A. Benjamin, New York, 1975).
\bibitem{Door09} P. Doornenbal et al., Phys. Rev. Lett. {\bf 103}, 032501
(2009).
\bibitem{kohama08} A. Kohama, K. Iida, K. Oyamatsu, Phys. Rev. C {\bf 78},
061601(R) (2008).
\bibitem{HS12} K. Hagino and H. Sagawa,Phys. Rev. C {\bf 84}, 011303(R) (2011);
Phys. Rev. C {\bf 85}, 014303 (2012).
\bibitem{minomo11} K. Minomo, T. Sumi, M. Kimura, K. Ogata, Y. R. Shimizu, and M. Yahiro, Phys. Rev. C {\bf 84}, 034602 (2011); 
T. Sumi, K. Minomo, S. Tagami, M. Kimura, T. Matsumoto, K. Ogata,
Y.R. Shimizu, and M. Yahiro, Phys. Rev. C {\bf 85}, 064613 (2012).
\bibitem{audi03} G. Audi, A. H. Wapstra, and C. Thibault, Nucl. Phys. {\bf A729},
337 (2003).
\bibitem{BFH87} P. Bonche, H. Flocard, and P.H. Heenen, Nucl. Phys. {\bf A467}, 115 (1987).
\bibitem{yordanov12} D. T. Yordanov {\it et al.}, Phys. Rev. Lett. {\bf 108},
042504 (2012).
\bibitem{neff12} T. Neff, private communication.
\bibitem{cameron12} J. Cameron, J. Chen, B. Singh, N. Nica,
Nuclear Data Sheets 113, 365 (2012).
\bibitem{takechi09} M. Takechi {\it et al.}, Phys. Rev. C {\bf 79}, 061601(R) (2009).
\bibitem{ozawa01} A. Ozawa {\it et al.}, Nucl. Phys. {\bf A693}, 32 (2001).
\bibitem{kanungo01} R. Kanungo, I. Tanihata, A. Ozawa,
Phys. Lett. {\bf B512}, 261 (2001).
\bibitem{kanungo02} R. Kanungo {\it et al.}, Phys. Rev. Lett. {\bf 88}, 142502 (2002).
\bibitem{ozawa00} A. Ozawa, T. Kobayashi, T. Suzuki, K. Yoshida, and I. Tanihata,
Phys. Rev. Lett. {\bf 84}, 5493 (2000).
\bibitem{kanungo11b} R. Kanungo {\it et al.}, Phys. Rev. C {\bf 84}, 061304 (R) (2011).
\end{thebibliography}
\end{document}